\documentclass[12pt]{article}
\newcommand{\al}{\alpha}

 \newcommand{\pr}{\prime}
 \newcommand{\ov}{\over}

\newcommand{\be}{\begin{equation}}
\newcommand{\ee}{\end{equation}} \newcommand{\bq}{\begin{eqnarray}}
\newcommand{\eq}{\end{eqnarray}} \newcommand{\nn}{\nonumber}
\newcommand{\ba}{\begin{array}} \newcommand{\ea}{\end{array}}

 \newcommand{\iy}{\infty} 
\newcommand{\la}{\lambda}  
\newcommand{\sg}{{\rm {sgn}}} \newcommand{\om}{\omega}
\newcommand{\Om}{\Omega} 
\newcommand{\tr}{\mbox{\rm tr\,}} 
 \newcommand{\br}{\bf R}
\newcommand{\de}{\delta} \newcommand{\ef}{\varepsilon_F}
 
   \newcommand{\hg}{{\hat g}}
 \renewcommand{\v}{{\rm v}}
\newcommand{\cg}{{\cal G}} 
\begin{document} \title{\bf The X-ray problem
revisited} \author{Estelle L. Basor\thanks{Supported in part by NSF Grants
DMS-9623278 and DMS-9970879.}\\ \emph {Department of Mathematics}\\ \emph
{California Polytechnic State University}\\ \emph {San Luis Obispo, CA
93407, USA} \and Yang Chen\\ \emph {Department of Mathematics}\\ \emph
{Imperial college}\\ \emph {180 Queen's Gate, London, SW7 2BZ, UK}} \date{}
\maketitle

\begin{abstract} In this letter we re-visit the X-ray problem. 
Assuming point interaction between the conduction electrons
and the instantaneously created core-hole, the latter's Green's
function can be represented as a Fredholm determinant of certain
Wiener-Hopf operators acting on $L^2(0,T)$ with discontinuous symbols. 
Here the symbols are the local conduction electron Green's function 
in the frequency domain and $T$ is the time the core-hole spend 
in the system before removal.  In this situation, the classical 
theory of singular integral equations usually employed in the 
literature to compute the large $T$ asymptotics of the Fredholm 
determinant ceased to be applicable.  A rigorous theory first 
put forward in the context of operator theory comes
into play and universal constants are found in the asymptotics. 
\end{abstract}

%%%%%%%%%%%%%%%%%%%%%%%%%%%%%%%%%%%%%%%%%%%%%%%%%%%%%%%%%%%%%%%%%%% 
We consider the classical X-ray problem where the core-hole is
created at the origin at $t=0$ and removed at $t=T>0.$ The object here is
to study the behaviour of the core-hole Green's function as $T$ gets large. 
By first integrating out the core-hole followed by the conduction electrons
\cite{Chen} or using a diagrammatic approach \cite{noz} the core-hole
Green's reads, 
\be \frac{\cg(T)}{\cg^{(0)}(T)}= \frac{\det(A-\v B)}{\det A},
\ee
where $\v>0,$ is the strength of local potential and $\cg^{(0)}$ is
the free core-hole Green's function.  

The operator $A$ has kernel, \be
\left(i\frac{\partial}{\partial t}+ \frac{1}{2m}\frac{\partial^2}{\partial
x^2}+\ef\right)\delta(x-x^{\prime}) \delta(t-t^{\prime}),\nonumber \ee and
$B$ is the multiplication operator \be \delta(x)\chi_{[0,T]}(t), \ee where
$m$ the mass of the conduction electrons, $\ef$ the Fermi energy, and for
simplicity we have assumeed that the one-dimensional electron 
gas has parabolic dispersion, $E_p=\frac{p^2}{2m}.$ A simple 
calculation shows that, \bq
\mbox{} -\ln\left(\frac{\cg(T)}{\cg^{(0)}(T)}\right)&=&
\tr\int_{0}^{\v}(A-\la B)^{-1}Bd\la\nn\\
\mbox{}&=&\int_{0}^{\v}\int_{0}^{T}G_{\la}(0,t;0,t+0)dtd\la\nn\\
\mbox{}&=:&\int_{0}^{\v}\int_{0}^{T}g_{\la}(t,t+0)dtd\la.  \eq Here
$G_{\la}$ is the time-ordered electron Green's function and satisfies \bq
\left(i\frac{\partial}{\partial t}+\frac{1}{2m} \frac{\partial^2}{\partial
x^2}+\ef-\la\de(x)\chi_{[0,T]}(t)\right)
G_{\la}(x,t;x^{\pr},t^{\pr})=\de(x-x^{\pr})\de(t-t^{\pr}), \eq or
equivalently the integral equation, 
\be
G_{\la}(x,t;x^{\pr},t^{\pr})=G_{0}(x,t;x^{\pr};t^{\pr})
+\la\int_{0}^{T}G_0(x,t;0,t^{\pr\pr})G_{\la}(0,t^{\pr\pr};x^{\pr},t^{\pr})
dt^{\pr\pr}.  \ee Putting $x=x^{\pr}=0$ in (6) gives the integral equation;
\be g_{\la}(t,t^{\pr})=g_0(t-t^{\pr})+\la\int_{0}^{T}g_0(t-t^{\pr\pr})
g_{\la}(t^{\pr\pr},t^{\pr})dt^{\pr\pr}, \ee where
$g_{\la}(t,t^{\pr}):=G_{\la}(0,t;0,t^{\pr})$.  

We define $F(\om)$ to be the inverse Fourier transform of $g_{0}(t)$ so 
that
\bq \mbox{} g_0(t)&=&\int_{-\infty}^{\infty}\exp(-i\om
t)F(\om)\frac{d\om}{2\pi}, \nn\\ \mbox{} F(\om)&=&\int_{-\infty}^{\infty}
\frac{\Om(\om)}{\om-\frac{p^2}{2m}+\ef+i0\sg\om}\frac{dp}{2\pi}, \eq 
where
$\Om$ is a Schwartz function which regulates the artificial ultraviolet
divergence due to the idealised point interaction between the core-hole and
the conduction electrons.  We suppose $\Om(0) = 1.$ It will be shown
later that the universal constant and exponents in $T$ are not affected by
this regularisation procedure.  
We find, by iterating (7), 
\bq
\frac{\cg(T)}{\cg^{(0)}(T)}=\det(I-\v\hg_{(0,T)}), \eq 
where \bq
\hg_{(0,T)}f(t):=\int_{0}^{T}g_0(t-t^{\pr})f(t^{\pr})dt^{\pr}.  \eq This is
the determinant representation of $\cg$ mentioned earlier and is the
starting point for our analysis of the large $T$ asymptotics.  In a
different physical context, to be described later the function $F,$ 
defined below will be
appropriately modified.  Computing the integral in (8), gives, 
\be
F(\om-\ef)=\Om(\om-\ef)\cases{i{\rm sgn}(\om-\ef){\sqrt {\frac{m}{2\om}}}
\;,& $0\leq\om<\infty$\cr -{\sqrt {\frac{m}{-2\om}}}\;,&$-\infty<\om<0$\cr}
\ee 
 In the classical theory of integral operators the function $F$ 
is called the symbol of the finite convolution operator, or finite 
Wiener-Hopf operator. The often 
stated theorem which yields asymptotics of determinants of such operators 
is the classical result of Kac-Akhiezer-Hirschman \cite{K,A, H1, H2}. 
However 
since the symbol has singularities at  $\om=\ef$ and  $\om=0$, this 
theorem no longer applies, and
alternative methods are required to deal with such singularities. We note 
here the theorem is already violated with one singularity.   
It is clear
that the kernel which is the Fourier transform of $F(\om-\ef)$ has
identical Fredholm determinant with that of $\det(I-\v\hg_{(0,T)}).$ 

There have been previous attempts in the physics literature to deal 
with such symbols, although much of the attempts are formal in nature.
In the
original paper \cite{noz}, where a flat band approximation is used,
$F(\om)$ is replaced by the following:
\be
F(\om)=D_0(\al+i\pi\sg\om),\;\;\om\in\br \ee here $D_0$ is the density of
states at the Fermi level and in this context $\ef$ is set to 0.  The
inverse Fourier transform of (12) gives rise to a linear combination
of $\delta(t)$ and $P\frac{1}{t},$ where $P$ is the principle value
operator.  Using this, (7) becomes a singular integral equation of the
form, 
\be
a\phi(t)+\frac{b}{i\pi}P\int_{0}^{T}\frac{\phi(t^{\pr})}{t-t^{\pr}}
dt^{\pr}=\varrho(t),\;\;t\in[0,T],\nonumber \ee where $a$ and $b$ are
constants and $a^2-b^2\neq 0.$ Now the ``driving term'' $\varrho,$ 
which is again a linear combination of a principal value operator 
and a Dirac delta, and which
is neither in $L^2[0,T]$ nor belongs to the H\"older class, invalidates a
straightforward application of the standard solution theory of singular
integral equations described for example in \cite{Mik}.  In \cite{noz}, a
formal computation, which claims to give an exact expression for
$\cg(T)/\cg^{(0)}(T),$ shows that it is asymptotic to 
\be
\exp\{a_1T+a_2\log T+a_3 +{\rm o}(1)\}, \ee as $T\to\infty,$ where
$a_j,\;j=1,2,3,$ are constants independent of $T.$ Indeed, $a_1,$ contains
a divergent term, which is, however, harmless, since it can be renormalised
by a redefinition of the core-level.  This is due to the choice of $F(\om)$
in (12).  There is also controversy \cite{Janis} as to the precise value of
$a_2.$ Finally $a_3$ has so far not been computed, heuristically or
otherwise, although attempts has been made \cite{oh}.  

 On the other hand, in the mathematics literature singular symbols have 
been studied for some time. Older results from this literature as well 
as some recent progress in evaluating 
such asymptotics rigourously confirm the form above 
of the asymptotics. The reader is referred to \cite{ba}, \cite{ba1}, 
\cite{bo}, \cite{bo1}, and \cite{BSW}.
We present those results here so that at least 
the value of the constants $a_{1}$ and $a_{2}$ can be conjectured 
with a high degree of certainty.

We begin with considering a class of symbols called the 
pure Fisher-Hartwig singular symbols. These are defined by
\[\psi_{\al,\beta}(\om)=\left({\om-0i\ov \om-i}\right)^{\al}
\left({\om+0i\ov \om+i}\right)^{\beta}.\]
(We specify the arguments of $\om\pm0i$ and $\om\pm i$ 
to be zero or close to it when $\om$ is
large and positive.)
This has the behavior
\[ 
\psi_{\al,\beta}(\om)\sim |\om|^{\al+\beta}\,e^{{1\ov2}i\pi(\al-\beta)
\,{\rm sgn}\om}\ \ \ {\rm as}\ \om\to0.\]
We associate a convolution operator to this symbol on $L^{2}[0,T]$ by 
convolving with the Fourier transform of $\psi -1.$ Let us call this 
operator plus the identity $W_{T}(\psi).$ Notice this symbol has the 
behavior of our previous $F$ defined in (11).

Since $\psi -1$ is in $L^{2}$ a determinant for $W_{T}(\psi)$ is not 
defined necessarily. We must use the regularized determinant 
$\mbox{det}_{2}W_{T}(\psi).$ In case both determinants are defined for
an operator $A$ we have
\[
\mbox{det}_{2}A = \det A\; {\rm e}^{-\mbox{tr}(A-I)}.
\]

Now it follows from the results in \cite{ba1} that if
$|{\cal Re}(\alpha \pm \beta)|<1$ then
\[\frac{\mbox{det}_{2} 
W_T(\psi_{\al,\beta})}{G_{2}(\psi_{\al,\beta})^{T}}\sim
 \left(\frac{T}{2}\right)^{\al\beta}E(\al,\beta)
\]
 where
\[G_{2}(\psi) = 
\exp\left(\frac{1}{2\pi}\int_{-\iy}^{\iy}(\log \psi(\om) 
-\psi(\om) +1)d\om\right),
\]
\[E(\al,\beta) = \frac{G(1 + \al )G(1 +\beta)}{G( 1 + \al + \beta)},
\]
and where $G$ is the Barnes G-function, an entire function satisfying
$G(z + 1) = \Gamma(z)G(z)$ with the initial condition $G(1) = 1.$

The results of \cite{ba1} can be modified so that a 
product of such functions times a smooth function can be considered.
This was already been done when $\al = -\beta$ and 
in a few other special cases. However, the more general case follows
now from the Toeplitz localization techniques of \cite{bo2} which can be modified
to apply to the Wiener-Hopf case as well. If we consider a product of two such 
functions, one with a singuarity at zero and one with a singuarity at 
$\ef$, and then times the product by a ``nice'' function $b$,
the asymptotic formula is as follows:
Let 
\[
\psi(\om) = b(\om)\, \psi_{\alpha_1,\beta_1}(\om) \psi_{\al_{2}, 
 \beta_{2}}(\om - \ef).
 \]
 Then
\begin{equation}
 \mbox{det}_{2}W_T(\psi)
\sim  G_{2}(\psi)^T \left(\frac{T}{2}
\right)^{\alpha_1\beta_{1} + \alpha_{2}\beta_{2}}E 
\end{equation}
where the constant $E$ is a product of three terms, that is 
$E = E_{1}E_{2}E_{3}$ where 
\[
E_{1} = \frac{G(1 + \al_{1} )G(1 +\beta_{1})}{G( 1 + \al_{1} + \beta_{1})}
\frac{G(1 + \al_{2} )G(1 +\beta_{2})}{G( 1 + \al_{2} + 
\beta_{2})}
\]and 
\[E_{2} = \exp\left(\frac{1}{2\pi}\int_{0}^{\infty}t S(t)S(-t)
dt\right)
\]
where $S(t)$ is the Fourier transform of $\log b(\om)$.
The last factor is given by
\[E_{3} = 
\frac{(1-i\ef)^{2\al_{1}\beta_{2}}(1+i\ef)^{2\al_{2}\beta_{1}}}
{(2i -\ef)^{\al_{2}\beta_{1}}(-2i -\ef)^{\al_{1}\beta_{2}}\ef^{\al_{2}\beta_{1}
+\al_{1}\beta_{2}}}
\]
\[\times
\frac{b_{-}(-i)^{\al_{1}}
b_{-}(\ef-i)^{\al_{2}}b_{+}(i)^{\beta_{1}}b_{+}(\ef+i)^{\beta_{2}}}
{b_{-}(0)^{\al_{1}}
b_{-}(\ef)^{\al_{2}}b_{+}(0)^{\beta_{1}}b_{+}(\ef)^{\beta_{2}}}.
\]
In this last factor all arguments are taken between $-\pi$ and $\pi$ 
and the functions $b_{\pm}$ are the normalized Wiener-Hopf factors of $b$.
For definitions of these factors and similar results the reader is 
referred to \cite{bo} where the constant was first conjectured, 
although in the above form appears here for the first time as far
as the authors are aware. Also, it should be noted that while
this is the fomula for the regularized determinant, in the case
that the ordinary determinant exists, say when $\al = \beta$ or if
the function $\Omega$ is sufficiently small, then the asymtotic 
formula given above in (15) only changes by the factor
$G_{2}$. It is replaced with 
$$G_{1}(\psi) = \exp(\frac{1}{2\pi}
\int_{-\iy}^{\iy}\log \psi(\om) d\om).$$ 
This follows immediately from the definition of the regularized determinant.

It is interesting to note that in the above the first factor is the
product of the constants that arise for a single singularity, the 
second term is the constant for nice symbols and the third factor
is what arises from the interaction of all three. 

Now we need to identity only the $\al$'s and $\beta$'s. This 
computation yields that $\al_{1}=0,$ 
$\beta_{1}=-1/2,$ $\al_{2}=-\beta_{2}=-\theta/\pi,$ where 
$\tan\theta=\v\Omega(0)\sqrt{\frac{m}{2\ef}}.$ This yields the following answer:
\begin{equation}
    \mbox{det}_{2}W_T(\psi)
\sim  G_{2}(1-\v F)^T \left(\frac{T}{2}
\right)^{-\frac{\theta^{2}}{\pi^{2}}}E_{1}E_{2}E_{3} 
\end{equation}
where
\[
E_{1} = 
G\left(1+\frac{\theta}{\pi}\right)G\left
(1-\frac{\theta}{\pi}\right)
\]and 
\[E_{2}=\exp\left(\frac{1}{2\pi}\int_{0}^{\infty}tS(t)S(-t)dt
\right)
\]
where $S(t)$ is the Fourier transform of 
$$\ln\left(\frac{1-\v F(\omega)}
{\psi_{0,-1/2}(\om) \psi_{-\theta/\pi , 
\theta/\pi}(\om - \ef)}\right).$$
The last factor is given by
\[
E_{3} = 
\frac{(1 +i\ef)^{\frac{\theta}{2\pi}}}
{(2i- \ef)^{\frac{\theta}{\pi}}\ef^{\frac{\theta}{2\pi}}}
\]
\[
\times
\frac{
b_{-}(\ef-i)^{\frac{-\theta}{\pi}}b_{+}(i)^{-1/2}b_{+}
(\ef+i)^{\frac{\theta}{\pi}}}
{b_{-}(\ef)^{-\frac{\theta}{\pi}}b_{+}(0)^{-1/2}b_{+}(\ef)^{\frac{\theta}{\pi}}}.
\]
%%%part2%%%%%%%
An allied problem which arises from the study of the X-ray
photo emission of a two-dimensional electron gas under the influence of a
constant magnetic field \cite{cal}, where \be
F(\om-\ef)=\Om(\om-\ef)\sum_{n=0}^{\infty} \frac{\exp(-\al
n)}{\om-n\om_{\rm c}+i\delta\sg(\om-\ef)}, \ee where $\al>0$ provides a high
energy cut-off and $\om_{\rm c},$ a fixed positive parameter, is the
cyclotron frequency and we think of $\delta$ as a small positive
constant which we later let tend to zero.

This symbol is again discontinuous and in this case
since $\al = -\beta$ the previously mentioned analysis also applies.
For this purpose, we first
compute the jump of $1-\v F$.  
We consider the quantity
\be 
\al - \beta = \frac{1}{i\pi} \ln\frac{(1-\v
F)_{\om=\ef^+}}{(1-\v F)_{\om=\ef^-}} 
\ee
This is seen to be
\be\frac{1}{i\pi} \ln (\frac{1- c +id
}{1 -c -id})
\ee where
\be
c = -v\Omega(0) \sum_{n=0}^{\infty} \frac{\exp(-\al
n)(\ef -n\om_{\rm c})}{(\ef-n\om_{\rm c})^{2}+\delta^{2}}
\ee
and 
\be
d = -v\Omega(0) \delta \sum_{n=0}^{\infty} \frac{\exp(-\al
n)}{(\ef-n\om_{\rm c})^{2}+\delta^{2}}
\ee
This yields asymptotics again of the form found in equation (15)
except that there is just one singularity. The form these take is
\begin{equation}
\mbox{det}_{2}W_T(\psi)
\sim  G_{2}(1-vF)^T \left(\frac{T}{2}
\right)^{-\theta^{2}/\pi^{2}}E
\end{equation}
where $\theta$ is the arctangent of $\frac{d}{1-c}.$ 

Let us now consider these asymptotics as $\delta $ tends to zero.
If $\Omega$ is nice enough, say bounded, continuous and integrable,
then the $G_{2}$ term tends to a limit. This is also true of the 
$\beta$ term. In fact, a straightforward calculation shows that
is $\ef \neq p\om_{\rm c}$ for any positive integer $p$ 
then $\theta$ tends to zero as $\delta$ tends to zero. 

If $\ef = p\om_{\rm c}$ for some $p$ then to compute the limit,
we notice that $d = a/\delta + O(\delta)$ with $a$ constant
and negative and that $1 - c = 1 + O(\delta).$ Thus as $\delta$
tends to zero, $\theta$ approaches $-\pi/2$ and the exponent 
in fromula (22) approached $-1/4$.

However, it should be pointed out that from the extensive work in the 
past that as far as the asymptotics go, these limits are sometimes 
correct at least in the first two terms, but that the constant terms
do not always yield a correct answer when limit processes are 
interchanged. In any case, if the X-ray case bears resemblence
to the classical work done in the Ising model, then the appearance
of the ``1/4'' is no surprise.

\end{document}